\documentclass[runningheads]{llncs}
\usepackage{amsmath}
\usepackage{amssymb}
\usepackage{placeins}
\usepackage[T1]{fontenc}
\usepackage{tabularx}
\usepackage{graphicx}
\usepackage{verbatim}
\usepackage{hyperref}
\usepackage{xcolor}
\DeclareUnicodeCharacter{03B3}{\ensuremath{\gamma}}
\definecolor{dark-red}{rgb}{0.4,0.15,0.15}
\definecolor{dark-blue}{rgb}{0.15,0.15,0.8}
\definecolor{medium-blue}{rgb}{0,0,0.5}
\hypersetup{
    colorlinks, linkcolor={dark-red},
    citecolor={dark-blue}, urlcolor={medium-blue}
}

\begin{document}
%

\title{Speed-Weighted Flocking for Sailing Swarms under Dynamic Environmental Forcing}
%
\titlerunning{Collective Behavior in Wind-Driven Robot Swarms}
%

\author{
Pranav Kedia\inst{1,2}\orcidID{0000-0003-4614-9540}
\thanks{Corresponding author}
\and
Aaron Gan\inst{3}
\and
Hannah J. Williams\inst{1,4,5}\orcidID{0000-0002-6338-529X}
\and
Andreagiovanni Reina\inst{1,2,5}\orcidID{0000-0003-4745-992X}
\and
Heiko Hamann\inst{1,2}\orcidID{0000-0002-2458-8289}
}

\authorrunning{P. Kedia et al.}

\institute{
Centre for the Advanced Study of Collective Behaviour,
University of Konstanz, Konstanz, Germany\\
\email{\{pranav.kedia, hannah.williams, andreagiovanni.reina, heiko.hamann\}@uni-konstanz.de}
\and
Department of Computer and Information Science,
University of Konstanz, Konstanz, Germany
\and
University of Pittsburgh, Pittsburgh, PA, USA
\and
Department of Biology,
University of Konstanz, Konstanz, Germany
\and
Max Planck Institute of Animal Behavior, Konstanz, Germany
}
\maketitle              
\begin{abstract}
Collective behavior models, such as aggregation and flocking, typically assume self-propelled robots can directly execute desired speeds and headings. 
However, autonomous sailing robots violate this assumption. Their motion is shaped by wind-dependent propulsion, restricted headings, and spatially varying wind conditions.
In particular, maneuverability is coupled to wind speed: in weak wind, sailboats may turn only slowly or not at all, whereas stronger wind enables faster turns. 
This introduces transient heterogeneity in speed and maneuverability across the flock. 
We focus on this fast-slow coordination problem in sailing robot flocks. 
To study this problem, we introduce \emph{SailSwarmSwIM}, a reduced-order simulator that captures wind-dependent speed and maneuverability, no-go zones, tacking behavior, and steady or gusty wind fields. 
To design our novel flocking technique, we start from the Couzin model and introduce a speed-weighted social interaction rule that accounts for each robot’s transient motion constraints. 
Across four wind conditions, moderate slow-neighbor weighting combined with speed matching under sailing constraints improves compactness, reduces unsafe events, and improves polarization under sailing constraints.
This effect arises from a balance between attraction to fast neighbors, which helps maintain movement, and cohesion around slow neighbors, which prevents the flock from fragmenting.
Together, SailSwarmSwIM and the speed-weighted interaction rule provide a framework for studying collective behavior in robotic fleets whose motion capabilities are continuously shaped by wind. 

\keywords{Autonomous Sailing Robots \and Sailing Swarm Simulator \and Flocking \and  Collective Behavior \and  Wind-powered Locomotion
}
\end{abstract}
\section{Introduction}

Classical models of collective motion show how coherent group-level behavior emerges from local interactions, such as repulsion, orientation, and attraction, without centralized control. 
Many foundational models of collective motion assume fully homogeneous system~\cite{Reynolds1987,Vicsek1995,Couzin2002}. Homogeneous social interactions and homogeneous locomotion capabilities are strong assumptions that can complicate the design of controllers for robots whose motion is constrained by actuation limits, morphology, and environmental forcing. 
More recent work has relaxed these assumptions by studying how individual heterogeneity shapes collective movement~\cite{MarDelgado18,jolles2020role}. 
Two-species Vicsek models~\cite{dutta2025stability} and~\cite{klamser21} show that variable individual speed and speed-dependent turning can qualitatively alter collective motion, affecting polarization, disorder transitions, spatial extent, and group-size effects. 

Our focus on motion-constrained heterogeneous collective motion originates from the challenge of building a swarm of autonomous sailing robots. 
Wind-driven sailing robots cannot move freely in arbitrary directions. 
Their speed and maneuverability depend on wind speed, wind direction, sail trim, rudder control, and their heading relative to the wind~\cite{Sun2022,Tipsuwan2023}. 
In weak winds, turns can become slow, unreliable, or even temporarily impossible; conversely, stronger winds can enable faster and more reliable maneuvers. 
Sailboats also cannot sail directly upwind but must perform tacking maneuvers to reach targets inside the upwind ``no-go'' zone~\cite{LeBars2015}. Any desired heading produced by a flocking controller must be transformed into a feasible sailing direction. The mapping from social interaction rules to realized motion is therefore indirect, state-dependent, and environmentally mediated.

These constraints become more important in a swarm. In a fleet of sailing robots, each robot may experience different yet spatially correlated local wind conditions, apparent wind angles, and tacking constraints~\cite{Sun2022}. 
Even when the robots are physically homogeneous, the environment introduces temporary heterogeneity in speed and maneuverability. 
Similar differences are common in animal collectives, where individuals exhibit phenotypic variation and may differ in speed, energetic state, age, morphology~\cite{harel21}, or maneuverability~\cite{jolles2020role}. 
Cohesion matters because a fragmented sailing swarm may lose local interaction links, leave wind-constrained robots behind, and cease to operate as a coherent sensing or monitoring team. We therefore treat reduced flock spread as improved cohesion, while evaluating unsafe proximity separately to avoid equating cohesion with crowding.
Maintaining cohesion often requires locomotor compromise through adaptive behaviors: faster or less constrained individuals may slow down, deviate, or otherwise adjust their movement to remain with slower or more constrained group members~\cite{sankey2019speed}. For artificial swarms, this raises an important design question: should social influence be weighted towards fast neighbors that can make rapid progress, or towards slower neighbors that risk being left behind?

Starting from a classical zonal flocking controller based on attraction, alignment, and repulsion~\cite{Couzin2002}, we introduce a speed-weighted social interaction rule. This rule defines a behavioral continuum ranging from following the fastest-moving neighbor to anchoring the flock around the slowest neighbors that are currently more constrained. 
Slow-neighbor weighting may improve cohesion and safety by preventing faster robots from overstretching or fragmenting the flock. A~central question is therefore whether an intermediate weighting regime can reduce proximity risks while preserving group polarization. 

Before implementing our flocking method on autonomous sailboat hardware, we test it in \emph{SailSwarmSwIM}, our dedicated reduced-order simulator for autonomous sailing robot swarms. 
Prototyping and testing in simulation is essential due to the high cost of deployment (e.g., unpredictable weather, high operational cost, and safety risks)~\cite{Sun2022,Tipsuwan2023}. This approach is also consistent with prior work on single-boat controllers, which are often tested in simulation before deployment~\cite{Vautier2018,Stenersen2016}. 
SailSwarmSwIM focuses on sailing-specific wind-dependent speed, no-go zones, tacking behavior, and configurable steady- or gusty-wind fields.

We formulate sailing robot flocks as collective-behavior systems shaped by wind-driven locomotor constraints. We compare the baseline Couzin-type controller~\cite{Couzin2002} against speed-weighted variants that combine modified social heading selection with sail luffing, a velocity-modulation mechanism that reduces propulsion by easing the sail. Our results show that moderate slow-neighbor weighting improves alignment, safety, and cohesion.

\section{Wind-Driven Sailing Robot Model}
\label{sec:model}

We use a reduced-order planar model. 
The model does not reproduce full six-degree-of-freedom hydrodynamics; instead, it captures the locomotor constraints relevant for swarm coordination: wind-dependent speed, infeasible upwind headings, tacking, and spatially or temporally varying wind. 
These constraints are sufficient to render the mapping from the desired social direction to the realized motion indirect and environment-dependent.

\paragraph{\textbf{Sailing terminology.}}
We use the following sailing terms throughout. The \emph{no-go zone} is the range of headings too close to the upwind direction for useful propulsion. 
\emph{Tacking} is the maneuver of alternating between feasible headings on either side of the wind to make upwind progress toward a target inside the no-go zone. 
A~\emph{close-hauled} heading is the closest feasible heading to the wind direction. 
\emph{Luffing}  reduces aerodynamic drive; we use it as an optional speed-matching mechanism (Sect.\,\ref{subsec:luffing}), separate from the speed-weighted social rule itself.

\paragraph{\textbf{Reduced-order sailing dynamics. }}

Each robot \(i\) has state
$\mathbf{x}_i = (x_i, y_i, \psi_i)$,
where \((x_i,y_i)\) is its position in the horizontal plane and \(\psi_i\) is its heading. 
The controller outputs a desired heading, which is mapped to feasible motion through simplified sailing dynamics. Forward speed depends on heading relative to local wind, reflecting that sailboats cannot directly command arbitrary velocities. Both sail trim and rudder action must be coordinated with the wind~\cite{Tipsuwan2023}. 
The position update is simply 
\begin{equation}
\begin{pmatrix}
x_i(t+\Delta t)\\
y_i(t+\Delta t)
\end{pmatrix}
=
\begin{pmatrix}
x_i(t)\\
y_i(t)
\end{pmatrix}
+
\Delta t
\left[
v_i(t)
\begin{pmatrix}
\cos \psi_i(t)\\
\sin \psi_i(t)
\end{pmatrix}
+
\mathbf{u}_{\mathrm{drift},i}(t)
\right],
\label{eq:position_update}
\end{equation}
for the wind-dependent forward speed~\(v_i(t)\), surface drift \(\mathbf{u}_{\mathrm{drift},i}(t)\), and simulation time step~$\Delta t$. 
We focus on wind-driven effects and use calm-water conditions. The heading updates by 
$\psi_i(t+\Delta t) = \psi_i(t) + \Delta t\,\omega_i(t)$,
where \(\omega_i(t)\) is the commanded yaw rate after applying sailing feasibility constraints.

\paragraph{\textbf{No-go constraint and tacking. }}

Headings inside an upwind no-go cone are infeasible and must be replaced. 
A~sailing robot can reach any waypoint virtually under any wind condition, but needs to tack. The robot does so by alternating between port and starboard headings to make upwind progress~\cite{LeBars2015}. 
When the desired direction lies within the no-go cone, the robot selects a nearby close-hauled heading and proceeds with tacking.  
Let \(\theta_w(\mathbf{x}_i,t)\) denote the local wind direction at robot \(i\), and let \(\alpha_{\mathrm{ng}}\) be the half-angle of the no-go cone. 
We define the signed angular difference between  desired heading and upwind direction as
\begin{equation}
\delta_i =
\operatorname{atan2}\left(
\sin(\psi_i^\star-\theta_w(\mathbf{x}_i,t)),
\cos(\psi_i^\star-\theta_w(\mathbf{x}_i,t))
\right).
\end{equation}
The desired heading is kept if \(|\delta_i| \geq \alpha_{\mathrm{ng}}\), and otherwise projected to the nearest close-hauled heading \(\tilde{\psi}_i = \theta_w(\mathbf{x}_i,t) + \operatorname{sgn}(\delta_i)\,\alpha_{\mathrm{ng}}\)
and \(\tilde{\psi}_i=\psi_i^\star\) otherwise.

This projection preserves the controller's intended direction as much as possible while enforcing the basic locomotor constraint of sailing. 
As a result, two robots receiving similar controller outputs may produce different motions depending on local wind conditions.

\paragraph{\textbf{Wind field. }}

The wind field is modeled as a base wind plus optional gusts:
$\mathbf{w}(x,y,t)
=
\mathbf{w}_{\mathrm{base}}(t)
+
\mathbf{w}_{\mathrm{gust}}(x,y,t)$. 
Here, \(\mathbf{w}(x,y,t)\in\mathbb{R}^2\) denotes the local wind-velocity vector.
Its magnitude \(\|\mathbf{w}(x,y,t)\|\) gives the local wind speed, and its
direction defines the local wind direction \(\theta_w(x,y,t)\). The base wind
$\mathbf{w}_{\mathrm{base}}$ sets the nominal speed and direction. Gusts are
modeled as a coherent sinusoidal disturbance that propagates along the
base-wind direction,
$
\mathbf{w}_{\mathrm{gust}}(\mathbf{p},t)
  = A\,\sin\!\big(\omega t - k\,(\hat{\mathbf{u}}\cdot\mathbf{p})\big)\,\hat{\mathbf{u}} ,
$
where \(k=2\pi/\lambda\) is the spatial
wavenumber, and \(\omega=k\,c\) is the temporal angular frequency induced
by propagation speed \(U_g\), $\mathbf{p}=(x,y)$. Here, $A$, $\lambda$, $c$ and $\hat{\mathbf{u}}$ denote amplitude, wavelength, propagation speed, and
base-wind direction.
In steady conditions, we set $A=0$; in gusty conditions we use $A = 2\,\mathrm{m/s}$, $\lambda = 50\,\mathrm{m}$, and $c = 10\,\mathrm{m/s}$ (temporal period $\lambda/c = 5\,\mathrm{s}$). 
These gusts introduce the spatial and temporal variation that induces transient
locomotor heterogeneity: mechanically identical robots differ in instantaneous
ability to follow the flocking rule depending on local wind, point of sail,
tacking requirements, and gusts. We evaluate four wind conditions: steady and
gusty variants of $5\,\mathrm{m/s}$ and $10\,\mathrm{m/s}$ base wind.

\paragraph{\textbf{Simulation loop. }}
At each step, the simulator evaluates local wind, computes the social heading, applies sailing-feasibility constraints, and advances the robot state using Eq.~\eqref{eq:position_update} and the heading update. 


\section{Flocking Controllers}
\label{sec:flocking_controllers}

We use a Couzin-type controller whose repulsion, orientation, and attraction zones represent collision avoidance, alignment, and cohesion. Speed-dependent influence is applied only to the social terms; short-range repulsion remains unchanged. The controller computes a desired social heading, which is passed through the sailing controller described in Sect.~\ref{sec:model}. We compare a baseline Couzin-type flocking controller with our novel speed-weighted variants.

\paragraph{\textbf{Baseline: Couzin-type flocking. }}
\label{subsec:baseline_couzin}

The baseline controller follows the zonal interactions of
Couzin-type flocking models~\cite{Couzin2002,Giardina2008}. Let \(\mathbf{p}_i=(x_i,y_i)^\top\) denote the planar position of robot $i$. 
Each robot \(i\) observes neighboring robots~$j\in \mathcal{N}_i$ within a finite interaction range and partitions them into three zones according to their distance \(d_{ij} = \|\mathbf{p}_j-\mathbf{p}_i\|\). 
Robots within the repulsion zone (\(d_{ij} \leq r_{\mathrm{rep}}\)) are too close and trigger collision-avoidance behavior.
Robots inside the orientation zone (\(r_{\mathrm{rep}} < d_{ij} \leq r_{\mathrm{ori}}\)) influence heading alignment.
Robots inside the attraction zone (\(r_{\mathrm{ori}} < d_{ij} \leq r_{\mathrm{att}}\)) pull toward the local group.
Here, \(r_{\mathrm{rep}}\), \(r_{\mathrm{ori}}\), and \(r_{\mathrm{att}}\) denote the radii of the repulsion, orientation, and attraction zones, respectively.
For each neighbor \(j\), we define $\hat{\mathbf{e}}_j=(\cos\psi_j,\sin\psi_j)^\top$. 
The repulsion, orientation, and attraction terms are then
\begin{equation}
\begin{aligned}
\mathbf{r}_i
&=
-\sum_{\substack{\scriptscriptstyle j\in\mathcal{N}_i\\ \scriptscriptstyle d_{ij}\le r_{\mathrm{rep}}}}
\frac{\mathbf{p}_j-\mathbf{p}_i}{d_{ij}},
&
\mathbf{o}_i
&=
\sum_{\substack{\scriptscriptstyle j\in\mathcal{N}_i\\ \scriptscriptstyle r_{\mathrm{rep}} < d_{ij}\leq r_{\mathrm{ori}}}}
\hat{\mathbf{e}}_j,
&
\mathbf{a}_i
&=
\sum_{\substack{\scriptscriptstyle j\in\mathcal{N}_i\\ \scriptscriptstyle r_{\mathrm{ori}}< d_{ij}\leq r_{\mathrm{att}}}}
\frac{\mathbf{p}_j-\mathbf{p}_i}{d_{ij}}.
\end{aligned}
\label{eq:couzin_terms}
\end{equation}
 
The controller uses the standard priority order:
repulsion, then orientation and attraction~\cite{Couzin2002,Giardina2008}.
If at least one neighbor lies inside the zone of repulsion, the desired social
direction is determined by \(\mathbf{r}_i\). Otherwise, the robot combines the
orientation and attraction terms $\mathbf{d}^{\mathrm{social}}_i
=
\mathbf{o}_i + \mathbf{a}_i$.
This direction-only Couzin controller is our baseline. 
It computes a desired social heading, while speed changes arise passively from wind-dependent sailing dynamics.

\paragraph{\textbf{Speed-weighted social influence. }}
\label{subsec:speed_weighted_flocking}

The baseline controller assigns equal influence to all neighbors within the same
interaction zone. In wind-driven robot flocks, this ignores the fact that robots
may differ in their instantaneous locomotor capacity. A~robot on a favorable point of sail may move and respond quickly, while an unfavorable apparent wind angle, a no-go constraint, or a tacking maneuver may slow another. 
We introduce a speed-weighted variant in which a neighbor's influence depends on their current speed. This information could be communicated between autonomous sailboats. 
For considered robot~$i$ and each of its neighboring robots~$j\in \mathcal{N}_i$, we define a regularized inverse-power law 
\begin{equation}
w_{ij}
=
\frac{1}{(v_j+\varepsilon)^\gamma}\;,
\label{eq:speed_weight}
\end{equation}
where $v_j>0$ is the current speed of neighbor~\(j\), \(\varepsilon\) is a small
regularization constant, and \(\gamma\) is the speed-weighting exponent. 
The weighted orientation and attraction terms are
\begin{equation}
\begin{aligned}
\mathbf{r}_{i}
&=
-\sum_{\substack{\scriptscriptstyle j\in\mathcal{N}_i\\\scriptscriptstyle d_{ij}\leq r_{\mathrm{rep}}}}
\frac{\mathbf{p}_j-\mathbf{p}_i}{d_{ij}},
&
\mathbf{o}_{\gamma,i}
&=
\sum_{\substack{\scriptscriptstyle j \in \mathcal{N}_i\\\scriptscriptstyle 
r_{\mathrm{rep}}< d_{ij}\leq r_{\mathrm{ori}}}}
w_{ij}\hat{\mathbf{e}}_j,
&
\mathbf{a}_{\gamma,i}
&=
\sum_{\substack{\scriptscriptstyle j \in \mathcal{N}_i\\\scriptscriptstyle 
r_{\mathrm{ori}}< d_{ij}\leq r_{\mathrm{att}}}}
w_{ij}\frac{\mathbf{p}_j-\mathbf{p}_i}{d_{ij}}
\end{aligned}
\label{eq:weighted_orientation_attraction}
\end{equation}
and the corresponding social vector is
$\mathbf{d}^{\mathrm{social}}_{\gamma,i}
=
\mathbf{o}_{\gamma,i}
+
\mathbf{a}_{\gamma,i}$. 

The exponent \(\gamma\) in Eq.\eqref{eq:speed_weight} is important as it defines the slow-fast behavioral axis of our novel flocking algorithm. 
For positive~$\gamma$, the rule gives larger weights to smaller~$v_j$, whereas for negative~$\gamma$ this preference is reversed. The offset~$\varepsilon>0$ removes the pole at~$v_j=0$, so the weighting remains finite over the admissible domain. 
We have several qualitatively different options: \(\gamma<0\) gives greater influence to faster neighbors, \(\gamma=0\) recovers uniform social weighting ($\forall{i,j\in \mathcal{N}_i}:w_{ij}=1$), and \(\gamma>0\) gives greater influence to slower neighbors. Importantly, \(\gamma=0\) does not remove wind-dependent speed variation from the robots; it only removes explicit speed-dependent social weighting. 
Here, \(\gamma\) changes the social weighting of neighbors rather than acting as a direct throttle command. 
The uniform Couzin baseline uses equal social weights and therefore does not exploit speed differences when determining the desired social direction. 
Sweeping \(\gamma\) allows us to find parameters balancing between cohesion, collision reduction, and polarization. The purpose is not to find a universally
optimal flocking controller, but to expose a slow-fast tradeoff. If wind-favored robots dominate the social direction, the flock may stretch or fragment as slower robots fall behind. If slow robots dominate, the group may remain compact but lose directional alignment. 

\paragraph{\textbf{Safety and boundary handling. }}

The repulsion term is not speed-weighted: local collision avoidance should not
depend on whether a neighbor is fast or slow. If a neighbor enters the repulsion
zone, the controller prioritizes separation using Eq.~\eqref{eq:couzin_terms}.
Although SailSwarmSwIM can accommodate COLREGs-based vessel right-of-way rules (international collision-avoidance rules for vessels at sea) for maritime encounters~\cite{Naeem2012}, we use a lightweight hard-repulsion layer to focus on flocking under wind-driven locomotor constraints. Candidate headings that would immediately violate the threshold $d_{\mathrm{rep}}=1.5 \,\mathrm{m}$ are rejected in favor of the nearest feasible alternative. 
While $d_{\mathrm{rep}}$ is used only to activate the hard-repulsion controller, 
safety is measured by the cumulative number of unsafe proximity events, defined
as pairwise distances below a smaller threshold $d_{\mathrm{near}}=1.0\, \mathrm{m}$. 

\paragraph{\textbf{Speed Equalization by Sail Luffing.}}
\label{subsec:luffing}
The speed-weighted rule sets heading but not speed, so a boat on a beam-reach may
outrun one sailing close-hauled or stalled near the no-go zone
($\beta_i \lesssim \alpha_{\mathrm{ng}}$). To reduce this heading-induced speed
dispersion, our speed-weighted controllers add a sail-luffing layer that depowers
fast boats toward the neighborhood speed. Sail trim is $\tau_i \in [0,1]$
(with $1$ fully powered). Luffing is asymmetric: reducing sail force
but not increasing it. 
With apparent-wind angle $\beta_i \in [0^\circ,180^\circ]$ sail easing is most effective near head-to-wind ($\beta_i \to 0^\circ$, lift-driven) and does little near dead
downwind ($\beta_i \to 180^\circ$, drag-driven). Therefore, the depower factor $D$ depends both on $\tau_i$ and $\beta_i$ as $D(\beta_i, \tau_i)= 1 - (1 - \beta_i/180^\circ)^{2}(1 - \tau_i)$. Each robot~$i$ sets its trim~$\tau_i$ by a rate-limited
proportional law driving it to the mean neighbor speed
$\bar v_i = \tfrac{1}{|\mathcal N_i|}\sum_{j \in \mathcal N_i} v_j$:
\begin{equation}
\tau_i \leftarrow \operatorname{clip}_{[0,1]}\!(
  \tau_i - \Delta \tau_i)\;,
\qquad \Delta\tau_i = \operatorname{clip}_{[-\Delta^\tau_\text{max},\Delta^\tau_\text{max}]} \Big( k_p\,\tfrac{v_i - \bar v_i}{\max(\bar v_i,\varepsilon_v)}\Big)\;.
\label{eq:trim}
\end{equation}
with proportional gain $k_p$, regularization $\varepsilon_v$, and per-step rate
limit $\Delta^\tau_\text{max}$.
A~boat faster than its neighbors eases and slows; one with no neighbors sets to $\tau_i = 1$. 
Since trim only depowers, the controller tends to reduce local speed dispersion and may approach the slowest locally sustainable pace.

Because speed determines social influence under the speed-weighting rule, luffing changes not only the physical motion of a boat but also its subsequent contribution to its neighbors' social decisions. The coupling is therefore bidirectional: speed-dependent weighting changes headings and, thereby, wind-induced speeds, while luffing changes speeds and, thereby, social weights. Consequently, the combined response is generated by a closed feedback loop rather than by two independent, additive mechanisms. Separate ablations remain possible, but they do not isolate context-independent contributions of weighting and luffing.

\section{Experimental Design and Statistical Analysis}
\label{sec:experiments}

All experiments use a square surface
arena with half-size \(L_{\mathrm{h}} = 35\,\mathrm{m}\), so that robot positions
lie in \(x,y \in [-L_{\mathrm{h}},L_{\mathrm{h}}]\). For flocking, we use a
toroidal interpretation of positions when computing neighbor relationships and
flock metrics, as in classical flocking studies. This avoids artificial
fragmentation or any wall effects when a coherent flock crosses an arena boundary, while the sailing dynamics and wind field are still evaluated in the square domain.

For each run, a swarm of \(N\) robots is initialized by sampling positions uniformly at random inside a central disk of radius \(R_0\), with headings sampled uniformly from \([-\pi,\pi)\). Unless otherwise stated, we use \(N=10\) robots and repeat
each controller--environment configuration over \(50\) random seeds. The
simulator advances with a time step of \(\Delta t = 1\,\mathrm{s}\) up to a
horizon of \(T=300\,\mathrm{s}\). 


\begin{table}[t]
\centering
\caption{Default simulation and controller parameters.}
\label{tab:default_params}
\setlength{\tabcolsep}{4pt}
\begin{tabular}{lll@{\hskip 12pt}lll}
\hline
param. & value & meaning & param. & value & meaning \\
\hline
\(N\) &
\(10\) &
number of robots &
\(\alpha_{\mathrm{ng}}\) &
\(45^\circ\) &
no-go half-angle \\

\(L_h\) &
\(35\,\mathrm{m}\) &
arena half-size &
\(r_{\mathrm{rep}}\) &
\(4\,\mathrm{m}\) &
social repulsion rad. \\

\(d_{\mathrm{rep}}\) &
\(1.5\,\mathrm{m}\) &
hard-repulsion threshold &
\(r_{\mathrm{ori}}\) &
\(10\,\mathrm{m}\) &
orientation radius \\

\(d_{\mathrm{near}}\) &
\(1.0\,\mathrm{m}\) &
unsafe-event threshold &
\(r_{\mathrm{att}}\) &
\(18\,\mathrm{m}\) &
attraction radius \\

\(\gamma\) &
\([-2,10]\) &
speed-weighting exponent &
\(k_p\) &
\(0.4\) &
luffing gain \\
\(\Delta^\tau_\text{max}\) & 0.05 & luffing trim rate limit \\
\hline
\end{tabular}
\end{table}

\paragraph{\textbf{Metrics.}} We evaluate three quantities: flock compactness, alignment, and safety. Flock
compactness is measured by the convex-hull area \(A_{\mathrm{hull}}(t)\) of the
robot positions. Lower area indicates a more compact flock. 
Alignment is measured by the polarization
\(\Phi(t)=\left\|N^{-1}\sum_{i=1}^{N}\hat{\mathbf{e}}_i(t)\right\|\),
where \(\Phi\approx1\) indicates aligned headings and \(\Phi\approx0\)
indicates disordered headings (under a maximum entropy assumption). 
This definition is consistent with the literature~\cite{Vicsek1995,Couzin2002,Giardina2008}.
Safety is measured by the cumulative number of unsafe proximity events recorded during the analysis window.


\paragraph{\textbf{Statistical analysis.}}For each independent simulation run, we ignore a transient phase ($t\in[0,100)$), and compute one steady-state
summary over the interval \(t\in[100,300]\,\mathrm{s}\). For flock area and
polarization, we use the median value over this interval. For unsafe proximity events, we use
the cumulative count. This produces one value per metric and simulation run.

We sweep \(\gamma\) over 25 values $\gamma\in [-2,10]$, with denser sampling near the exponent sign change at \(\gamma=0\) to resolve the transition between fast-neighbor following, uniform Couzin weighting, and slow-neighbor anchoring. We evaluate each controller under four wind conditions (each under a calm-water condition): \(5\,\mathrm{m/s}\) steady wind, \(10\,\mathrm{m/s}\) steady wind, \(5\,\mathrm{m/s}\) wind with gusts, and \(10\,\mathrm{m/s}\) wind with gusts.

Each speed-weighted setting is compared against the baseline Couzin controller
using seed-matched paired differences
$\Delta_{\gamma,s}
=
M_{\gamma,s}
-
M_{\mathrm{base},s}$, 
where \(M_{\gamma,s}\) is the metric value for tested exponent~\(\gamma\) and seed~\(s\), and \(M_{\mathrm{base},s}\) is the corresponding baseline value for the
same seed and environment. For flock area and unsafe proximity events, negative
\(\Delta_{\gamma,s}\) indicates improvement (smaller metric values are better). For polarization, positive
\(\Delta_{\gamma,s}\) indicates improvement (higher alignment is better).
Because unsafe-proximity events, are discrete and skewed, and because the paired
differences for some metrics show tail deviations from normality, we use
Wilcoxon signed-rank tests for pairwise comparisons against the baseline. For each metric and environment, \(p\)-values are corrected across all tested \(\gamma\) values using Holm correction. 
We report a change as a significant improvement only if Holm-corrected $p<0.05$ and the median paired difference is in the improvement direction.

\section{Results}
\label{sec:results}

\begin{figure*}[t]
    \centering
    \includegraphics[width=0.91\textwidth]{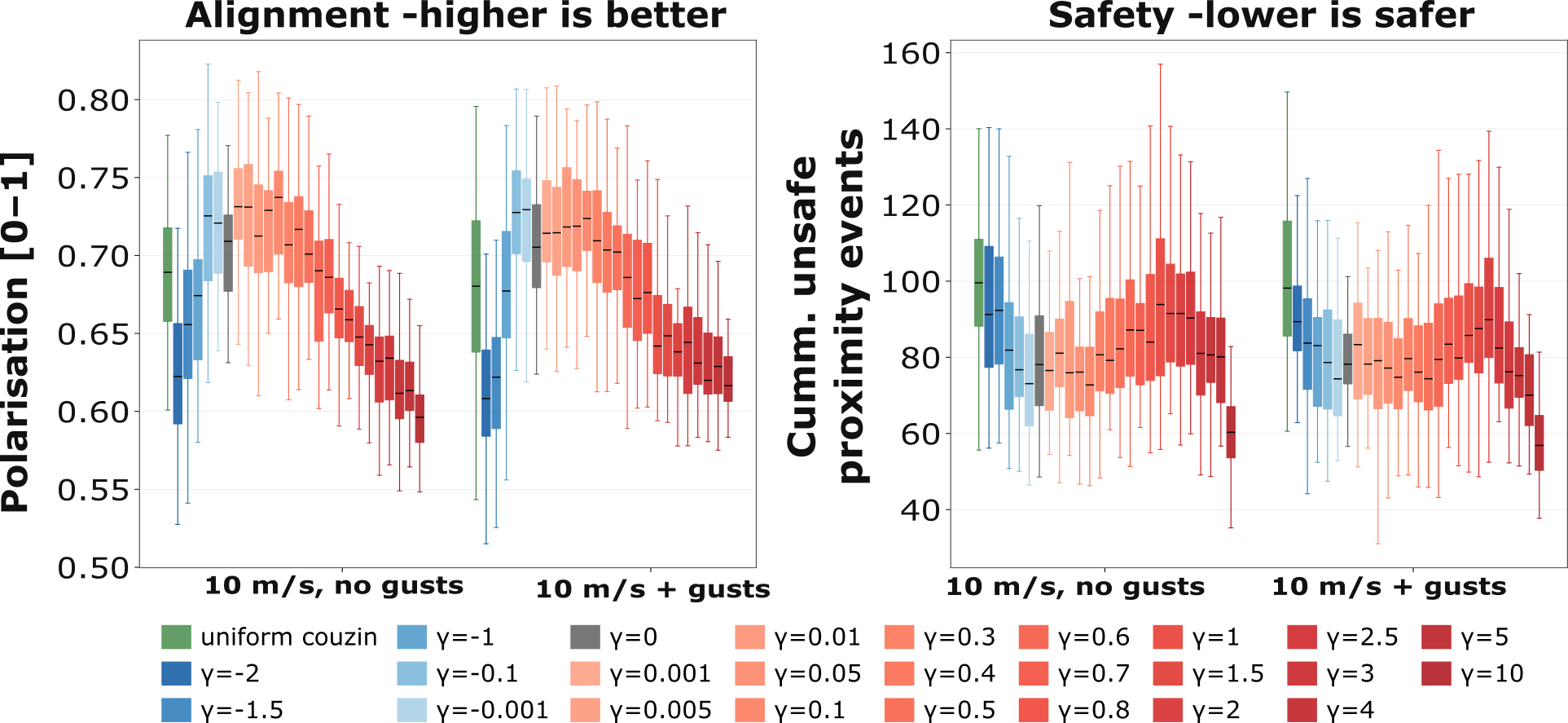}
    \caption{Steady-state alignment and safety at \(10\,\mathrm{m/s}\) across 50 runs per \(\gamma\). Higher polarization and fewer cumulative unsafe-proximity events indicate better performance.}
    \label{fig:perseed_overview}
\end{figure*}

\begin{figure}[t]
    \centering
    \includegraphics[width=0.96\textwidth]{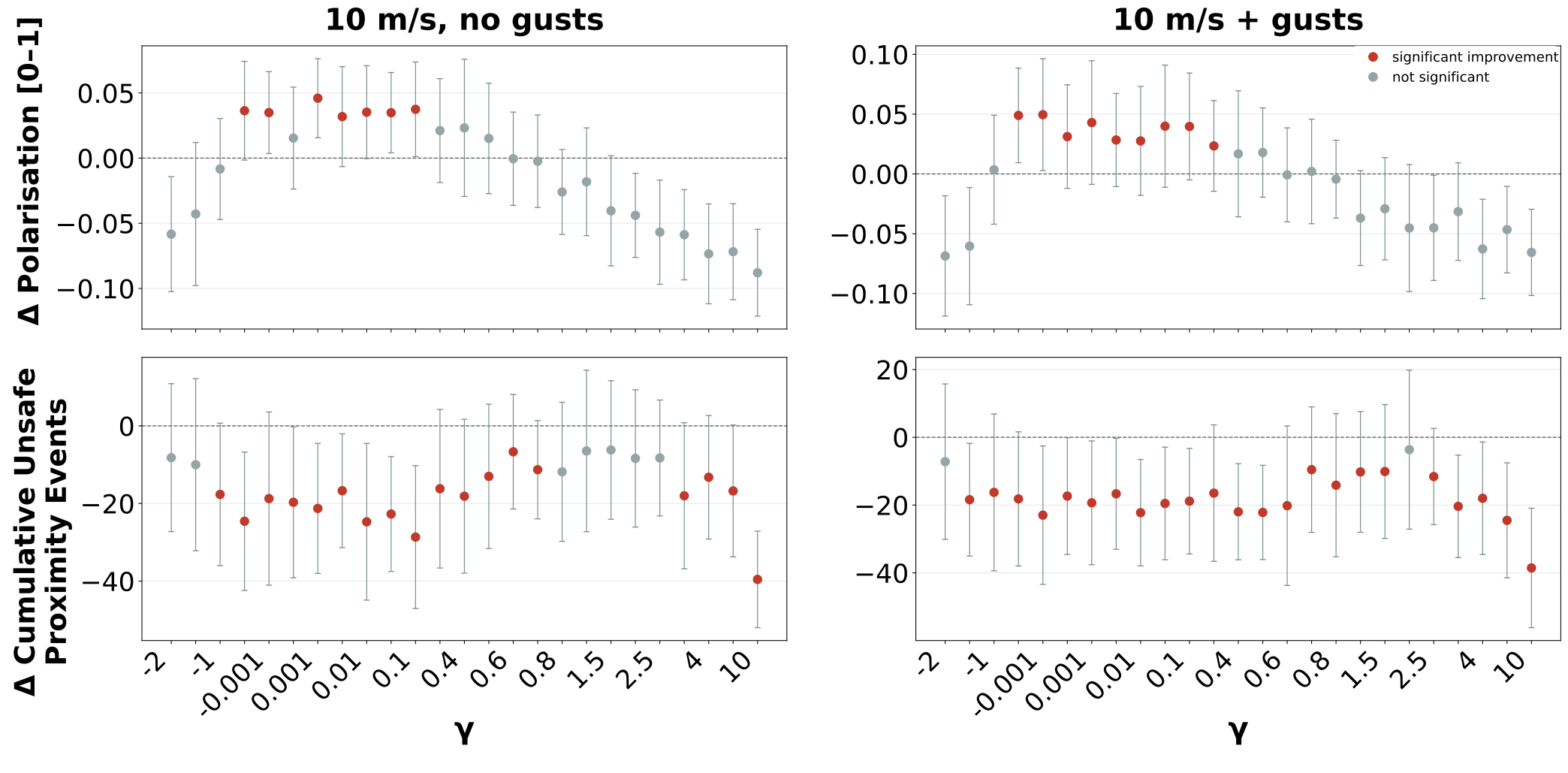}
    \caption{Effect of speed weighting at \(10\,\mathrm{m/s}\): median paired differences from the uniform Couzin baseline over 50 seed-matched runs, with interquartile ranges. Positive \(\Delta\Phi\) and negative \(\Delta C\) indicate improved alignment and safety, resp.}
    
    \label{fig:gamma_sweep_deltas}
\end{figure}

Fig.~\ref{fig:perseed_overview} shows the results for alignment and safety for the two $10\,\mathrm{m/s}$-wind settings based on our sweep of the speed-weighting exponent~\(\gamma\) (Eq.~\ref{eq:speed_weight}). 
Statistical significance and improvements over the baseline are given in Fig.~\ref{fig:gamma_sweep_deltas} and Table~\ref{tab:gamma_summary}. Results in the other conditions (5\,m/s wind) and for the compactness metric are available in the online supplementary material,\footnote{
\url{https://praked.github.io/publications/SailSwarmSim-SAB2026}
} along with code and videos.

\noindent\textbf{Moderate Slow-Neighbor Weighting under Luffing Provides the Best Compromise. }The exponent~\(\gamma\) parameterizes a continuous social-influence axis: \(\gamma\!<\!0\) gives faster neighbors greater authority over heading (fast-neighbor following) and \(\gamma\!>\!0\) anchors social influence toward slower neighbors. 

Alignment exhibits a clear inverted-U shape, with a central region above our Couzin model baseline. 
Safety also shows clear improvements over the baseline with a relatively complex system behavior for~$\gamma>1$. 
Within a narrow band \(\gamma\!\in\![-0.1,0.3]\) every cell of Fig.~\ref{fig:gamma_sweep_deltas} registers a significant improvement over the baseline, without any significant degradations. 
Outside this band, the picture deteriorates asymmetrically. Strong fast-neighbor following (\(\gamma\!\leq\!-1.5\)) significantly stretches the flock and reduces
polarization: in the \(10\,\mathrm{m/s}\) environments, at
\(\gamma=-2\), median flock area increases by
\(121.6\) and \(136.9\,\mathrm{m}^{2}\) under steady and gusty
wind, respectively, and polarization drops by~0.04--0.07. 
Strong slow-neighbor anchoring (\(\gamma\!\geq\!1\))
preserves or improves cohesion but progressively erodes alignment,
culminating at \(\gamma=10\) with a polarization loss of \(0.09\) and
an area increase of \(194.7\,\mathrm{m}^{2}\) under steady
\(10\,\mathrm{m/s}\) wind. The two failure modes are visible as opposite limbs of Fig.~\ref{fig:gamma_sweep_deltas}.

\noindent\textbf{A Robust Compromise across Cohesion, Safety, and Alignment. }Within the central \(\gamma\!\in\![-0.1,0.3]\) plateau for safety, small positive exponents~\(\gamma\) provide significant gains in both safety and also cohesion as seen in  Table~\ref{tab:gamma_summary} for the representative case~\(\gamma=0.01\). 
The convex flock area shrinks by
\(45.6\,\mathrm{m}^{2}\) to \(106.2\,\mathrm{m}^{2}\), corresponding to a
\(15\) to \(29\%\) reduction relative to the baseline median
(\(272\,\mathrm{m}^{2}\) to \(376\,\mathrm{m}^{2}\)). 
Cumulative unsafe proximity events drop by between~\(13.6\) and~\(24.7\) events, which is a \(22\) to $32\%$ reduction relative to the baseline. 
Polarization rises modestly by \(0.028\) to \(0.080\). 
The improvement is largest in the highest-energy environment (\(10\,\mathrm{m/s}\) with
gusts).

Within this coupled feedback loop, two effects plausibly contribute to the joint improvement in cohesion–safety. 
First, increased weighting of slower neighbors pulls the flock's centroid toward stalled or windward robots. 
Hence, we do not observe long-tail stretching following uniform alignment in gusty wind.
Second, the sail-luffing layer reduces the relative speeds between leaders and trailers, lowering approach speed and the rate at which the unsafe-proximity layer is triggered; removing luffing erases most of this safety gain and can push unsafe events back to baseline levels (Fig.~S10).
Comparing the behaviors based only on the luffing layer ($\gamma=0$) with those incorporating both speed-weighting and luffing ($\gamma \neq 0$; see Figs.\,S4--S10), shows that speed weighting significantly increases polarization but has no significant effect on the other metrics.

\begin{table}[t]
\centering
\caption{
Effect of the slow-anchoring setting \(\gamma=0.01\) relative
to the baseline (paired differences,
50 seed-matched runs; brackets show Cohen's \(d_z\)). Negative values improve
area and unsafe proximity events; positive values improve polarization. Asterisks denote significance (\(^{*}\,p<0.05\); \(^{**}\,p<10^{-3}\); \(^{***}\,p<10^{-5}\)).
}
\label{tab:gamma_summary}
\begin{tabular}{lrrrr}
\hline
Environment & \(\gamma\) &
\(\Delta A_{\mathrm{hull}}\,[\mathrm{m}^{2}]\) &
\(\Delta \Phi\) &
\(\Delta C\) \\
\hline
\(5\,\mathrm{m/s}\), steady  & \(0.01\) &
  \(-61.4^{***}\,[-0.72]\)  &
  \(+0.080^{***}\,[+0.99]\) &
  \(-16.9^{***}\,[-0.73]\) \\
\(10\,\mathrm{m/s}\), steady & \(0.01\) &
  \(-93.8^{***}\,[-0.93]\)  &
  \(+0.035^{*}\,[+0.49]\)   &
  \(-24.7^{**}\,[-0.64]\)  \\
\(5\,\mathrm{m/s}\), gusts   & \(0.01\) &
  \(-45.6^{*}\,[-0.52]\)    &
  \(+0.052^{*}\,[+0.53]\)   &
  \(-13.6^{***}\,[-0.68]\) \\
\(10\,\mathrm{m/s}\), gusts  & \(0.01\) &
  \(-106.2^{***}\,[-0.98]\) &
  \(+0.028^{**}\,[+0.60]\)  &
  \(-22.2^{***}\,[-0.75]\) \\
\hline
\end{tabular}
\end{table}
In summary, we find that the speed-weighting layer must be combined with luffing to improve flocking performance, and that \(\gamma=0.01\) provides a robust default across metrics, but not as an optimum for any single metric. 
Within the plateau, per-metric optima differ: in the \(10\,\mathrm{m/s}\) environments, area is minimized near \(\gamma\!\approx\!1\) and unsafe proximity events near \(\gamma=10\), but both settings significantly reduce polarization. 
By contrast, \(\gamma=0.01\) sacrifices a few percent of per-metric improvement in exchange for simultaneous gains on all metrics across all environments, without significant degradation. 

\begin{figure}[t]
    \centering
    \includegraphics[width=0.842\textwidth]{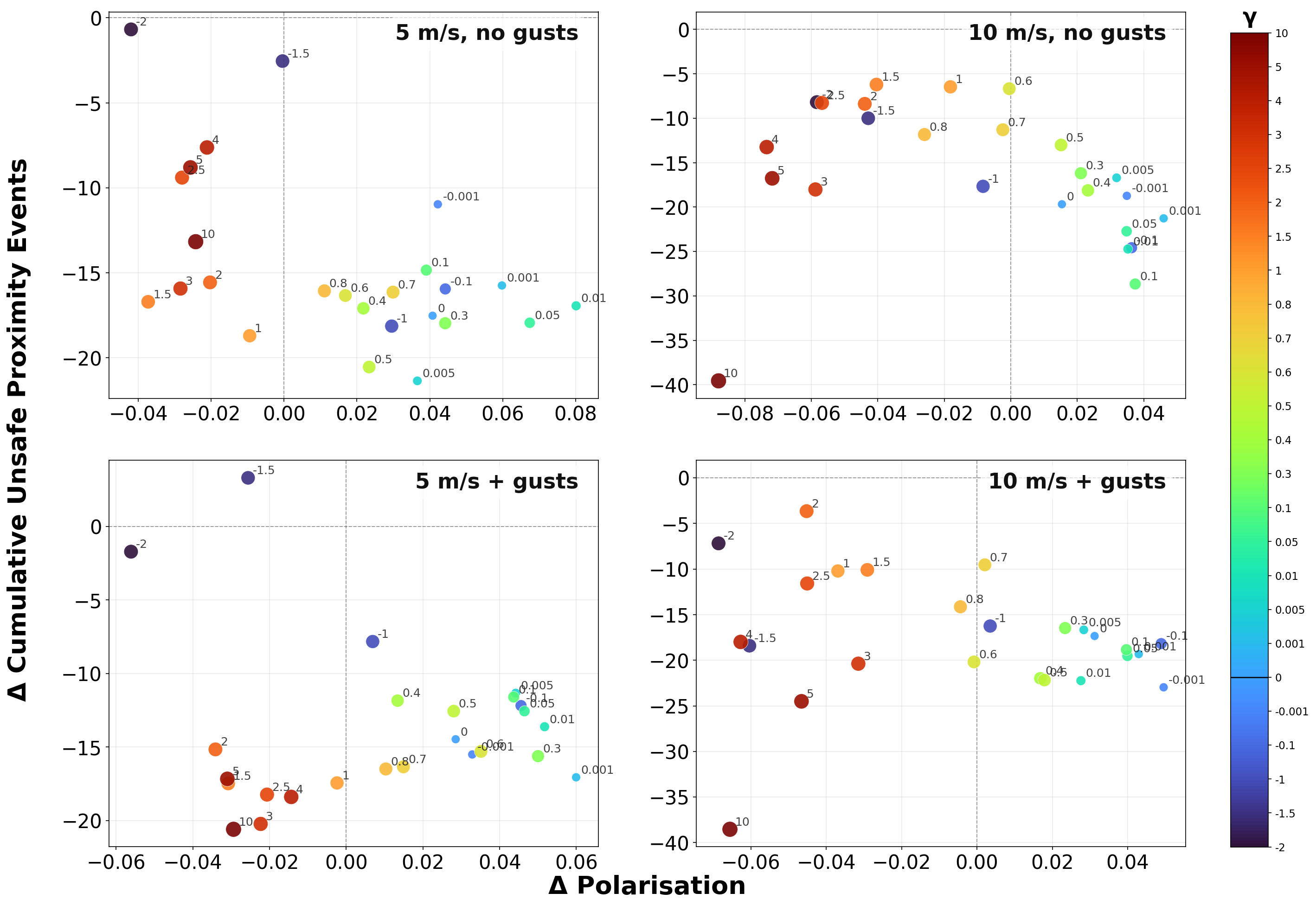}
    \caption{
    Alignment--safety tradeoff across wind conditions. Markers are \(\gamma\) settings as a median paired effect relative to the uniform Couzin
    baseline. The favorable region is lower right: higher polarization and fewer
    unsafe proximity events. 
    }
    \label{fig:tradeoff_plots}
\end{figure}

\noindent\textbf{Over-Anchoring Reveals an Alignment Tradeoff. }Pushing \(\gamma\) beyond the plateau exposes a sharp alignment tradeoff.
For \(\gamma \geq 1\) in the \(10\,\mathrm{m/s}\) environments, alignment drops significantly relative to baseline. 
Under steady wind, \(\Delta\Phi\) decreases from
\(-0.018\) at \(\gamma=1\) to \(-0.088\) at \(\gamma=10\), with qualitatively the same monotone degradation under gusts. 
Cohesion continues to improve over this range (\(\Delta A_{\mathrm{hull}}\) remains negative until \(\gamma=10\), at which point it reverses to \(+194.7\,\mathrm{m}^{2}\) under steady \(10\,\mathrm{m/s}\) wind). 
This can be seen as over-anchoring: an overly strong influence of slow robots preserves cohesion but degrades alignment. 
In Fig.~\ref{fig:tradeoff_plots}, we show this tradeoff by plotting changes in polarization against changes in unsafe proximity events. 
The desirable region is the lower-right quadrant (increased polarization and reduced unsafe proximity). 
Across environments, small positive~$\gamma$ settings occupy this region. 


\section{Conclusion}
\label{sec:conclusion}
We studied flocking in wind-driven robot swarms where social interaction rules
are filtered through sailing-specific locomotor constraints. Because no-go zones,
tacking, and local wind conditions make some robots temporarily faster or more
constrained than others, flocking becomes a slow-fast coordination problem. 
This study shows the effect of speed-weighted social influence under luffing using $N=10$ robots, calm-water conditions, and toroidal boundary handling for flock metrics. 
Safety is represented by a simplified hard-repulsion layer, and robots are assumed to have accurate access to neighboring positions, headings, and speeds. 
In physical sailing robotic swarms, these estimates may be affected by limited sensing and communication range, localization noise, delay, and packet loss. 
We introduced a speed-weighted Couzin controller with luffing in which \(\gamma<0\) implements fast-neighbor following and \(\gamma>0\) implements
slow-neighbor anchoring. Across four wind conditions, moderate slow-neighbor anchoring improved compactness, increased safety, and preserved or improved polarization. 
Our analysis shows that luffing conditionally contributes to improvements in safety and cohesion, while the speed-weighting exponent governs the alignment–cohesion tradeoff across the sweep. Stronger anchoring revealed an alignment tradeoff. Future work should therefore test goal-related tasks, highly temporally variable wind conditions, non-toroidal environments, and robustness to imperfect neighbor-state information.
\begin{credits}
\subsubsection{\ackname}
This work has been supported by the DFG under Germany's Excellence
Strategy, EXC 2117-422037984.

\subsubsection{\discintname}
The authors have no competing interests to declare.
\end{credits}

\bibliography{references}

\end{document}